\documentclass[12pt]{ronbun}

\catcode`\@=11
\@addtoreset{equation}{section}

\global\arraycolsep=2pt
\usepackage{amsbsy,amssymb,latexsym,amsfonts,amsmath,cite,epsfig}

\newcommand\CL{{\mathcal L}}

\newcommand\CH{{\mathcal H}}
\renewcommand\mod{~\mathrm{mod}~}

\renewcommand\NG{\mathrm{NG}}
\newcommand\WZ{\mathrm{WZ}}
\newcommand\e{\mathrm{e}}

\begin{document}

%\hfill
%KEK-TH-1135 \\
%\hfill
%OIQP-07-01
\begin{flushright}
\parbox{4.2cm}
{KEK-TH-1135 \hfill \\ 
 OIQP-07-01 \\ 
{\tt hep-th/0702062}
 }
\end{flushright}

\vspace{7mm}

\begin{center}
{\Large \bf
Intersecting Noncommutative M5-branes \\
from Covariant Open Supermembrane
}
\end{center}
\vspace{7mm}

\centerline{\large Makoto Sakaguchi$^{a}$ and Kentaroh Yoshida$^{b}$}
\vspace{7mm}

\begin{center}
$^a$ {\it Okayama Institute for Quantum Physics\\
1-9-1 Kyoyama, Okayama 700-0015, Japan}\\
makoto$\_$sakaguchi@pref.okayama.jp
\vspace{5mm}

$^b${\it Theory Division, Institute of Particle and Nuclear Studies, \\ 
High Energy Accelerator Research 
Organization (KEK),\\
1-1 Oho, Tsukuba, Ibaraki 305-0801, Japan.} 
\\
kyoshida@post.kek.jp
\end{center}

\vspace*{1cm}

\begin{abstract}
We study intersecting noncommutative (NC) M5-branes from
$\kappa$-invariance of an open supermembrane action with constant
three-form fluxes. The $\kappa$-invariance gives rise to possible
D-brane configurations for which projection operators can be determined.
We construct projection operators for two types of 1/4 BPS intersecting
NC M5-branes. The one is an intersection of two NC M5-branes: NC
M5$\bot$NC M5 (3). The other is that of a NC M5-brane and a commutative
(C) M5-brane: NC M5$\bot$C M5 (1). A NC M5-brane can be viewed as a
bound state of M5 and M2, and the configurations M2$\bot$M5 (1) and
M2$\bot$M2 (0) are realized on the intersecting M5-branes. Taking a
commutative limit the allowed intersecting M5-branes are surely
reproduced: M5$\bot$M5 (3) and M5$\bot$M5 (1).
\end{abstract}

\thispagestyle{empty}
\setcounter{page}{0}

\newpage

\section{Introduction}

Supermembrane theory in eleven dimensions \cite{BST,dWHN} is closely
related to the M-theory formulation \cite{BFSS}. Membranes (M2-branes)
are the fundamental objects and open membranes
\cite{Strominger,Townsend} as well as closed ones can be considered.
Open membranes can end on a $p$-dimensional hypersurface (Dirichlet
brane) with $p=1,5$ and 9 \cite{EMM,dWPP} just like an open string can
attach to D-branes. The $p=5$ case corresponds to M5-brane and the $p=9$
is the end-of-world 9-brane appearing in the Horava-Witten theory
\cite{HW}. Open M5-branes have been discussed in the recent work
\cite{openM5}.

\medskip
 
The Dirichlet branes can be investigated from the $\kappa$-symmetry
argument \cite{EMM}\footnote{M5-branes can be discussed from the
superembedding method \cite{CS}.}. This method is also applicable to
string theory \cite{LW}. It is a covariant way and a specific
gauge-fixing such as light-cone gauge is not necessary. Then it is
sufficient to consider a single action of open string or open membrane,
rather than each of D-brane actions. It is moreover easy to find what
configurations are allowed to exist for rather complicated D-brane
setups such as intersecting D-branes or less supersymmetric D-branes,
which are difficult to discuss within 
%%%
a brane probe analysis.
% approximation. 
Finally the
method is not restricted to a flat spacetime and can be generalized to
curved backgrounds\footnote{For Dirichlet branes of open supermembrane
in AdS$_{4/7}\times$S$^{7/4}$ and pp-wave see \cite{AdS-memb}. For
applications to D-branes in AdS$_5\times$S$^5$ and pp-wave, see
\cite{various}. For the results of the brane probe, see \cite{SMT}.}.
  
\medskip 

In this paper we discuss an application of the $\kappa$-symmetry
argument to supersymmetric intersecting M-branes
\cite{PT,Tseytlin,GKT1}.  There is, however, an obvious obstacle that an
open supermembrane can attach to M5-branes but not M2-branes due to the
charge conservation law \cite{Strominger}. Thus it is an easy task to
find intersecting M5-branes, but it would be more involved to consider
intersecting configurations including M2-branes. A possible way to
discuss M2 within the framework of this method is to consider an
M5-brane with 
%%%
electric and magnetic fluxes, 
which is called
noncommutative (NC) M5-brane \cite{SW,Bergshoeff}. It can be viewed as a
bound state of M5 and M2, and hence one may find a configuration of
intersecting M-branes with M2 on the intersecting M5-branes.

\medskip 

By following our previous paper \cite{SY:NCM}, we discuss two types of
1/4 BPS intersecting NC M5-branes: NC M5$\bot$NC M5 (3) and NC M5$\bot$C
M5 (1). The allowed configurations M2$\bot$M2 (0) \cite{PT,GKT1} and
M2$\bot$M5 (1) \cite{PT,GKT1} can be found on the intersecting
M5-branes. The supergravity solution corresponding to NC M5$\bot$NC M5
(3) has been constructed \cite{Kim}. These are the possible
configurations of intersecting M-branes (For a review of intersecting
D-branes and M-branes without fluxes, see \cite{Gauntlett:1997cv}). 
Taking a commutative limit leads to 
the possible configuration of intersecting M5-branes: M5$\bot$M5 (3)
\cite{PT,Tseytlin,GKT1} and M5$\bot$M5 \cite{GKT1}.  

\medskip \medskip

This paper is organized as follows. In section 2 we introduce the
covariant Green-Schwarz (GS) action of an open supermembrane in flat
space with constant three-form and derive the surface terms coming from
the $\kappa$-variation of the Wess-Zumino term in the action. In section
3 we elaborate NC M5-branes and strong flux limit of them. This section
is basically an exposition of \cite{SY:NCM}. In section 4 two types of
intersecting NC M5-brane configurations are constructed. The
configurations, which include M2-branes, M2$\bot$M5 (1) and M2$\bot$M2
(0) are found on the intersecting M5-branes. The projection operators
for them are found by considering a strong flux limit and we can check
that the surface terms from the $\kappa$-variation surely
vanish. Section 5 is devoted to a summary and discussions.

\section{Open supermembrane and $\kappa$-symmetry}

The GS action of a supermembrane in flat spacetime is composed of the
Nambu-Goto (NG) part and the Wess-Zumino (WZ) part \cite{BST}
\begin{eqnarray}
S&=&\int_\Sigma d^3\sigma\left[
\CL_{\NG}
+\CL_{\WZ}
\right]\,. \label{total}
\end{eqnarray}
The NG part is given by 
\begin{eqnarray}
&& \CL_{\NG} =
-\sqrt{-g(X,\theta)}\,,  \quad 
g_{ij} 
=E^A_iE^B_j\eta_{AB}\,, \quad 
E^A_i= \partial_iX^A -i\bar{\theta}\Gamma^A\partial_i\theta\,. 
\nonumber 
\end{eqnarray}
The WZ part is given by 
\begin{eqnarray}
\CL_{\WZ}&=&
\epsilon^{ijk}
\Big[
-\frac{1}{6}e^{A}_ie^{B}_je^{C}_k\CH_{ABC}
+\frac{i}{2}\bar\theta\Gamma_{AB}\partial_i\theta~
 \partial_jX^Me^A_M~
 \partial_kX^Ne^B_N
 \nonumber\\&&
+\frac{1}{2}\bar\theta\Gamma_{AB}\partial_i\theta~
\bar\theta\Gamma^A\partial_j\theta~
\partial_kX^Me^B_M
-\frac{i}{6}\bar\theta\Gamma_{AB}\partial_i\theta~
\bar\theta\Gamma^A\partial_j\theta~
\bar\theta\Gamma^B\partial_k\theta
\Big]\,. \nonumber 
\end{eqnarray} 
This part includes a coupling term to a three-form field 
$\CH =C-db$\,, where $C$ and $b$ are a three-form gauge potential 
and a two-form gauge potential on the brane, respectively. 

\medskip 

The total action (\ref{total}) is invariant under the $\kappa$-variation 
\begin{eqnarray}
\delta_\kappa X^{M}e^A_M=-i\bar\theta\Gamma^A\delta_\kappa\theta\,. \nonumber
\end{eqnarray}
This local fermionic symmetry ensures 
%%%
the consistency 
of the theory. This
symmetry is obviously maintained considering a {\it closed}
supermembrane. However our current interest is an {\it open}
supermembrane and for the action to be $\kappa$-symmetric we need to
impose some appropriate boundary conditions. These conditions should
describe Dirichlet branes of open supermembrane.

\medskip 

First of all, let us see boundary conditions for the bosonic coordinates
\cite{Bergshoeff}. Suppose a $p$-dimensional hypersurface as a Dirichlet
brane of an open supermembrane. When a constant $\CH$ is turned on along
the Dirichlet $p$-brane worldvolume $\Sigma$\,, they should satisfy
either of the following boundary conditions at a boundary of the world
volume $\partial\Sigma$\,,
\begin{eqnarray}
  \begin{array}{lll}
\partial_n X^{\bar {A}}
+\CH^{\bar A}{}_{\bar B\bar C}\partial_\tau X^{\bar {B}}\partial_tX^{\bar {C}}
=0\,,       
&& \quad \bar{A}_a (a=0,\cdots,p) \in \mbox{Neumann}  \\
\partial_\tau X^{\underline{A}}=\partial_tX^{\underline{A}}=0\,, 
&& \quad \underline{A}_a (a=p+1,\cdots,10) \in \mbox{Dirichlet}    \\
  \end{array}
\,, \nonumber 
%\label{boundary condition}
\end{eqnarray}
where $n$ is a normal direction to $\partial\Sigma$\,, and $\tau$ and
$t$ denote tangential ones. In a large $\CH$ limit the Neumann
directions for which fluxes are turned on are frozen and replaced by
Dirichlet ones\footnote{In the large $\CH$ limit, say $\CH_{\bar A_0\bar
A_1\bar A_2}\to\infty$, the Neumann directions $\{\bar A_0, \bar
A_1,\bar A_2\}$ reduce into $ \{\bar A_0\in D \cup \bar A_1\in D\} \cap
\{\bar A_1\in D \cup \bar A_2\in D\} \cap \{\bar A_2\in D \cup \bar
A_0\in D\} $, where $\bar A_0\in D$ denotes that $\bar A_0$ is
Dirichlet. For example, this is satisfied $\{\bar A_1\in D \cap \bar
A_2\in D\}$\,. Thus taking this limit a few Neumann directions are
frozen and the value $p$ reduces to $(p-2)$ or $(p-3)$\,.}.

\medskip  

Then let us discuss boundary conditions for the fermionic variables
$\theta$\,, which is sensitive to how much supersymmetries can be
maintained.  For the Dirichlet brane to be supersymmetric, ``gluing
conditions'' should be imposed on $\theta$ by constructing projection
operators. The main subject is how to construct the projection operators
for supersymmetric Dirichlet branes. The projection operators can be
determined from the requirement that the surface terms coming from the
$\kappa$-variation should vanish. No surface term appears when a
membrane is closed. However we are now considering an open membrane and
hence surface terms may appear and must be deleted in order to ensure
the consistency of the theory.

\medskip 

Since the bulk
%%%
action admits $\kappa$-symmetry,
% is assumed, 
the $\kappa$-variation of
the action $\delta_\kappa S$ leaves only surface terms. The NG part does
not give rise to any surface terms. Thus it is sufficient to examine the
$\kappa$-variation of the WZ part, 
\begin{eqnarray}
\delta_\kappa S_{\WZ}|&=&\int_{\partial\Sigma} d ^2\xi
\left[
\CL^{(2)}+\CL^{(4)}+\CL^{(6)}
\right]\,, \nonumber \\
\CL^{(2)}&=&-i
\left[
\bar\theta\Gamma_{\bar A\bar B}\delta_\kappa\theta
+\CH_{\bar A\bar B\bar C}\bar\theta\Gamma^{\bar C}\delta_\kappa\theta
\right]\dot X^{\bar{A}}{X'}^{\bar B}~,
\label{second order in theta}
\\
\CL^{(4)}&=&
\Big[
-\frac{3}{2}\bar\theta\Gamma^{ A}\delta_\kappa\theta~
\bar\theta\Gamma_{ A\bar B}
+\frac{1}{2}\bar\theta\Gamma_{ A\bar B}\delta_\kappa\theta~
\bar\theta\Gamma^{ A}
\Big](\theta'\dot X^{\bar B}-\dot \theta {X'}^{\bar B})\,, 
\label{surface term theta^4}
\\
\CL^{(6)}&=&
\frac{i}{6}
\Big[
\bar\theta\Gamma_{  A  B}\dot\theta~
\bar\theta\Gamma^{  A}\theta'~
\bar\theta\Gamma^{  B}\delta_\kappa\theta
-
\bar\theta\Gamma_{  A  B}\theta'~
\bar\theta\Gamma^{  A}\dot\theta~
\bar\theta\Gamma^{  B}\delta_\kappa\theta
\nonumber\\&&~~~
-2\bar\theta\Gamma_{  A  B}\delta_\kappa\theta~
\bar\theta\Gamma^{  A}\dot\theta~
\bar\theta\Gamma^{  B}\theta'
\Big]\,, 
\label{surface term theta^6}
\end{eqnarray}
where 
%%%
$\dot Z=\partial_\tau Z$ and $Z'=\partial_t Z$\,.
$\CL^{(n)}$ represents the term with the $n$-th order of $\theta$\,. 

\medskip 

It has been shown in \cite{SY:NCM} that $\CL^{(6)}$ in (\ref{surface
term theta^6}) vanishes due to the Fierz identity
\begin{eqnarray}
(C\Gamma_{AB})_{(\alpha\beta}(C\Gamma^A)_{\gamma\delta)}=0\,. \label{Fierz}
\end{eqnarray}
For the configuration that satisfies that $\CL^{(2)}=0$\,, the surface
terms in (\ref{surface term theta^4}) are rewritten with (\ref{Fierz})
as
\begin{eqnarray}
\CL^{(4)}=-\frac{1}{2}\left[
\bar\theta\Gamma^{\underline A}\delta_\kappa\theta~
\bar\theta\Gamma_{\underline A\bar B}
+
\bar\theta\Gamma_{\underline A\bar B}\delta_\kappa\theta~
\bar\theta\Gamma^{\underline A}
\right]
(\theta'\dot X^{\bar B}-\dot \theta{X'}^{\bar B})\,. 
\label{surface term theta^4 arranged}
\end{eqnarray}
Therefore the problem of finding possible Dirichlet branes is boiled
down to constructing the projection operators to make (\ref{second order
in theta}) and (\ref{surface term theta^4 arranged}) vanish.

\medskip 

In the next section we will construct projection operators for NC
M5-branes. 

\section{NC M5-brane and strong flux limit} 

Let us elaborate NC M5-branes as Dirichlet branes 
of an open supermembrane. This section is an exposition of 
\cite{SY:NCM}, but more careful derivation will be presented. 
It would also be helpful to make the manuscript self-contained.

\medskip 

The construction is slightly modified in the presence of the flux. We
shall first remember the case without flux for simplicity. Then we
construct projection operators for a NC M5-brane. It is regarded as a
bound state of M5 and M2 and in a strong flux limit, infinitely many
M2-branes are dissolved on the M5-brane.

\subsection{Dirichlet branes without flux \label{3.1}}

Let us concentrate here on 1/2 BPS Dirichlet branes
without flux.  By using a gluing matrix $M$, which consists of a product
of gamma matrices, the gluing condition is written as 
\begin{eqnarray}
\theta=M\theta\,, \quad 
M=\ell\Gamma^{\bar A_0\bar A_1\cdots\bar A_p}\,, \quad 
\ell^2(-1)^{[\frac{p+1}{2}]}s=1\,. 
\label{Mp projection}
\end{eqnarray}
Here $s=-1$ when $0\in\{\bar A_0,\bar A_1,\cdots,\bar A_p\}$ and $s=1$
otherwise.  It can be easily seen that $M$ satisfies the following
relations, 
\begin{eqnarray}
M^2=1\,, \quad 
\bar\theta=\bar\theta M'\,, \quad 
M'=(-1)^{p+1+[\frac{p+1}{2}]}M\,. \nonumber 
\end{eqnarray}

\medskip 

First of all, we find a condition under which (\ref{second order in
theta}) should vanish.  It is an easy task to show the following
identities,
\begin{align}
\bar\theta\Gamma_{\bar A\bar B}\delta_\kappa\theta
&=\frac{1}{2}\bar\theta(M'\Gamma_{\bar A\bar B}+\Gamma_{\bar A\bar B}M)\delta_\kappa\theta
=0 
&&\text{for $p=1,4\mod 4$}\,, 
\nonumber \\
\bar\theta\Gamma^{\bar C}\delta_\kappa\theta
&=0 
&&\text{for $p=3,4\mod 4$}\,.
\nonumber 
\end{align}
Hence (\ref{second order in theta}) vanishes when $p=1,4\mod 4$,
$\CH\equiv 0$\,.

Then let us consider a condition to delete (\ref{surface term theta^4
arranged}). Noting that
\begin{align}
\bar\theta\Gamma_{ \underline{A}\bar B}\delta_\kappa\theta
&
=0
&&\text{for $p=2,3\mod 4$}\,, \nonumber 
\\
\bar\theta\Gamma^{\underline{C}}\delta_\kappa\theta
&=0
&&\text{for $p=1,2\mod 4$}\,,   
\label{eqn 4}
\end{align} 
one can see that (\ref{surface term theta^4 arranged}) vanishes for
$p=1\mod 4$\,. Thus we have reproduced the classification of the 1/2
BPS Dirichlet branes in flat spacetime without flux \cite{EMM}.

\subsection{NC M5-branes}

Next we shall consider a NC M5-brane by including constant fluxes. Then
we need to generalize the ansatz for the gluing matrix as
follows\footnote{ The gluing matrix $M=h_0\Gamma^{\bar A_0\bar A_1\bar
A_2} +h_1\Gamma^{\bar A_3\bar A_4\bar A_5}$ leads to the same results
\cite{SY:NCM}.  This provides the same NC M5-brane with a different
parametrization.} 
\begin{align}&
M=h_0\Gamma^{\bar A_0\bar A_1\cdots\bar A_5}
+h_1\Gamma^{\bar A_0\bar A_1\bar A_2}~.
\label{M-A}
\end{align}
It is
%%%
characteristic of NC branes
that $M$ is represented by a 
{\it sum} of the products of gamma matrices in comparison to the case
without flux (\ref{Mp projection}). 

For the gluing matrix $M$ to define a projection operator, the condition
$M^2=1$ should be satisfied. Then we obtain the following condition, 
\begin{eqnarray}
-s_0h_0^2-s_1h_1^2=1\,. 
\label{M5+M2: M^2=1}
\end{eqnarray}
Here $s_0=-1$ when $0\in \{\bar A_0,\bar A_1,\cdots,\bar A_5\}$ and
$s_0=1$ otherwise, and $s_1=-1$ when $0\in \{\bar A_0,\bar A_1,\bar
A_2\}$ and $s_1=1$ otherwise. It would be helpful to see that the matrix
$M$ satisfies
\begin{eqnarray}
\bar\theta=\bar\theta M'\,, \quad 
M'=-h_0\Gamma^{\bar A_0\bar A_1\cdots\bar A_5}
+h_1\Gamma^{\bar A_0\bar A_1\bar A_2}\,. \nonumber 
\end{eqnarray}

\medskip 

Let us first examine (\ref{second order in theta}). Since we can easily
show the following identities,
\begin{eqnarray}
&&\bar\theta\Gamma_{\bar A_0\bar A_1}\delta_\kappa\theta
=\frac{1}{2}\bar\theta(M'\Gamma_{\bar A_0\bar A_1}+\Gamma_{\bar A_0\bar A_1}M)\delta_\kappa\theta
=-h_1\bar\theta\Gamma^{\bar A_2}\delta_\kappa\theta\,, \nonumber \\
&&\bar\theta\Gamma_{\bar A_3\bar A_4}\delta_\kappa\theta
=h_1\bar\theta\Gamma^{\bar A_0\bar A_1\bar A_2}{}_{\bar A_3\bar A_4}\delta_\kappa\theta\,, \qquad \bar\theta\Gamma_{\bar A_2\bar A_3}\delta_\kappa\theta=0\,,
\nonumber \\
&&\CH_{\bar A_3\bar A_4\bar A_5}\bar\theta\Gamma^{\bar A_5}\delta_\kappa\theta
=-h_0\CH^{\bar A_3\bar A_4\bar A_5}\bar\theta\Gamma^{\bar A_0\bar A_1\bar A_2}{}_{\bar A_3\bar A_4}\delta_\kappa\theta\,, \nonumber 
\end{eqnarray}
we can see that (\ref{second order in theta}) may vanish by imposing the
conditions
\begin{eqnarray}
h_1-\CH_{\bar A_0\bar A_1\bar A_2}=0~,~~~
h_1-h_0\CH^{\bar A_3\bar A_4\bar A_5}=0~.
\label{solution M5}
\end{eqnarray}
With (\ref{eqn 4}),  (\ref{surface term theta^4 arranged}) also becomes
zero. Therefore the gluing matrix (\ref{M-A}) with 
the two conditions (\ref{M5+M2: M^2=1}) and 
(\ref{solution M5}) gives a possible M5-brane configuration. 

\medskip 

Then let us consider the interpretation of the solution constructed
above.  For reality of $\CH$\,, it is sufficient to consider
$(s_0,s_1)=(-1,\pm 1)$\,. By substituting (\ref{solution M5}) for
(\ref{M5+M2: M^2=1}), we obtain the following condition,
\begin{eqnarray}
\frac{1}{(\CH^{\bar A_3\bar A_4\bar A_5})^2}
-\frac{1}{(\CH_{\bar A_0\bar A_1\bar A_2})^2}
=s_1\,. \nonumber 
%\label{SD M5}
\end{eqnarray}
This is nothing but the self-dual condition \cite{Sezgin} of the gauge
field on the M5-brane \cite{SW}. That is, we have reproduced the
information on the NC M5-brane from the $\kappa$-symmetry argument for
the open supermembrane action. Thus we recognize that the projection
operator should describe a NC M5-brane ($\bar A_0\cdots\bar A_5$) with
$\CH_{\bar A_0\bar A_1\bar A_2}$ and $\CH^{\bar A_3\bar A_4\bar
A_5}$\,. It is worth noting that the corresponding supergravity solution
is found in \cite{Kim}.

\subsubsection*{Commutative and strong flux limits}

Now let us examine a commutative limit and a large $\CH$ limit of the NC
M5-brane. 

\medskip 

We first consider the case with $s_1=-1$\,, say  NC M5 (012345) with
$\CH_{012}$ and $\CH^{345}$.
The conditions (\ref{M5+M2: M^2=1}) and (\ref{solution M5})  are solved
by using an angle variable $\varphi~(0\le \varphi \le\pi/2)$\,, 
\begin{eqnarray}
&&h_0=\cos\varphi\,, \quad h_1=\sin\varphi\,, \quad 
\CH_{012}=\sin\varphi\,, \quad 
\CH^{345}=\tan\varphi\,. 
\label{s1}
\end{eqnarray}
With (\ref{s1})\,, we can express the gluing matrix $M$ as 
\begin{eqnarray}
M=\e^{\varphi\Gamma^{345}}\Gamma^{012345}\,. \nonumber 
\end{eqnarray}
For a commutative limit $\varphi\to 0$, the NC M5 reduces to commutative
M5 (012345), since $\CH\to0$ and $M\to \Gamma^{012345}$.

\begin{figure}[ht]
  \begin{center}
    \includegraphics[keepaspectratio=true,height=50mm]{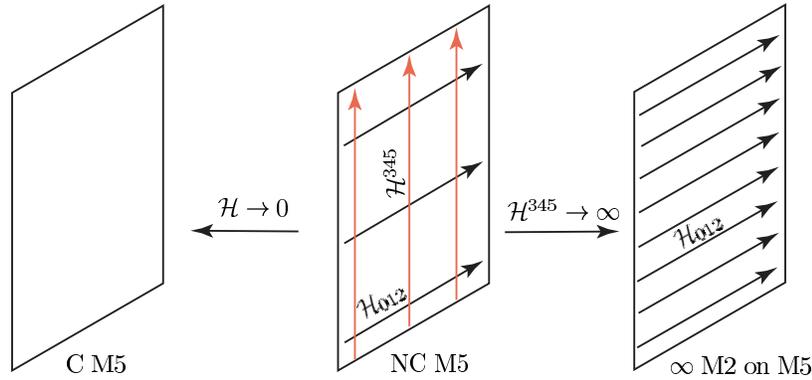}
  \end{center}
  \caption{The commutative and strong flux limits of NC M5-brane.}
  \label{fig:NCM5.eps}
\end{figure}

On the other hand, for $\varphi\to\pi/2$\,, we see that
$\CH^{345}\to\infty$ and so the gluing condition reduces to
$M\to\Gamma^{012}$ with a critical flux $\CH_{012}=1$\,. It seems that
the resulting projection operator should describe a critical M2-brane
(012). Eventually this limit is nothing but the OM limit \cite{OM} and
it should correspond to one of infinitely many M2-branes dissolved on
the M5-brane.  This is analogous to the D2-D0 setup where a D2-brane
with a flux reduces to a D2-brane with infinitely many D0-brane in a
strong magnetic flux limit. We summarize the results in Fig.\
\ref{fig:NCM5.eps}.

\medskip 

Next we examine the case with $s_1=1$, say the NC M5 (012345) with
$\CH_{345}$ and $\CH^{012}$.  The conditions (\ref{M5+M2: M^2=1}) and
(\ref{solution M5}) are solved again by using a single variable
$\varphi~(0\le\varphi<\infty)$
\begin{eqnarray}
&& h_0=-\cosh\varphi\,, \quad 
h_1=\sinh\varphi\,, \quad 
\CH_{345}=\sinh\varphi\,, \quad 
\CH^{012}=-\tanh\varphi\,. \nonumber 
%\label{s2} 
\end{eqnarray}
In this case the range of $\varphi$ is not bounded. 
Then the gluing matrix can be expressed as 
\begin{eqnarray}
M=\cosh\varphi\Gamma^{012345}+
\sinh\varphi\Gamma^{345}
=\e^{\varphi\Gamma^{012}}\Gamma^{01234 5}\,.  \nonumber 
%\label{A-2}
\end{eqnarray} 
In a commutative limit $\varphi\to 0$\,, the NC M5 reduces to a
commutative M5 (012345), since $h_0=-1$\,, $h_1=0$ and
$\CH_{345}=\CH^{012}=0$\,.  

On the other hand, to discuss a strong flux limit $\varphi\to\infty$, we
should note that the boundary condition for $\theta$ can be rewritten as
\begin{eqnarray}
2\e^{-\varphi}\theta
=[(1+\e^{-2\varphi})\Gamma^{012345}+
(1-\e^{-2\varphi})\Gamma^{012}\Gamma^{012345}]\theta~. \nonumber 
\end{eqnarray}
Then, after taking this limit, it turns to 
\[
0=\Gamma^{012345}(1-\Gamma^{012})\theta
\]
and so we obtain that 
\begin{eqnarray}
\theta=\Gamma^{012}\theta\,.
\end{eqnarray}
The physical interpretation of the resulting M2-brane with a critical flux 
$\CH^{012}=-1$ is the same as in the case with $s_1=-1$\,.

\vspace*{0.5cm}

\vspace*{0.5cm}

\subsubsection*{$\kappa$-invariance for a critical M2-brane}

When a strong flux limit is taken, the gluing condition for a NC
M5-brane reduces to that for one of the infinitely many M2-branes. As we
have already seen in Sec.\ \ref{3.1}, the $p=2$ case is not allowed as a
projection operator in the case without fluxes. Therefore it is a
non-trivial problem whether the resulting projection operator for a
critical M2 is consistent to the $\kappa$-symmetry,
%%%
or equivalently
%   In other words, it is 
whether the projection should delete the surface terms coming from
the $\kappa$-variation.

\medskip 

The $p=2$ case is actually special among other $p$\,, and we can
find the identities intrinsic to $p=2$\,,  
\begin{eqnarray}
\CH_{\bar A_0\bar A_1\bar A_2}\bar\theta\Gamma^{\bar A_2}\delta_\kappa\theta
=\CH_{\bar A_0\bar A_1\bar A_2}\bar\theta\Gamma^{\bar A_2}\ell\Gamma^{\bar A_0\bar A_1\bar A_2}
\delta_\kappa\theta
=\CH^{\bar A_0\bar A_1\bar A_2}\ell\bar\theta\Gamma_{\bar A_0\bar A_1}\delta_\kappa\theta \nonumber 
\end{eqnarray}
and so (\ref{second order in theta}) vanishes when
\begin{eqnarray}
1+\ell\CH^{\bar A_0\bar A_1\bar A_2}=0~.
\label{fixed H for NC M2 }
\end{eqnarray}
The flux $\CH$ should be real so that $\ell$ is real and thus $s=-1$\,.
It follows from (\ref{eqn 4}) that (\ref{surface term theta^4 arranged})
disappears. Thus we have checked that the $\kappa$-variation surface
terms should vanish for an M2-brane with a critical $\CH$ fixed by
(\ref{fixed H for NC M2 })\,.

Although the $\kappa$-symmetry is maintained for the M2-brane, the
charge conservation requires the existence of M5-brane behind
M2-branes. That is, a NC M5-brane should be regarded as a bound state of
M5 and M2.

\section{Intersecting NC M5-branes}

The main subject of this paper is to construct a projection operators
for intersecting NC M5-branes. A single NC M5-brane is characterized by
the product of gamma matrices with an exponential factor, as we have
already seen. Hence let us consider to describe intersecting NC
M5-branes by introducing the two gluing matrices, 
\begin{eqnarray}
M_1=\e^{\varphi_1\Gamma^{A_0A_1A_2}}\Gamma^{A_0\cdots A_5}\,, \quad
M_2=\e^{\varphi_2\Gamma^{B_0B_1B_2}}\Gamma^{B_0\cdots B_5}\,,\quad 
[M_1,M_2]=0\,. 
\label{gluing}
\end{eqnarray}
In order to avoid an imaginary $\CH$\,, the conditions
$0\in\{A_0,\cdots,A_5\}$ and $0\in\{B_0,\cdots,B_5\}$ are assumed. 
%We are interested in 1/4 BPS configurations and hence 
%the two gluing matrices should commute each other,
%\[
% [M_1,M_2]=0\,. 
%\] 

\medskip 

In the case without flux, i.e., $\varphi_1=\varphi_2=0$, the matrices
commute each other, if the number, say $n$, of the common indices
contained in $\Gamma^{A_0\cdots A_5}$ and $\Gamma^{B_0\cdots B_5}$ is
even. For $n=4$ it is an intersection of M5-branes along a 3-brane,
M5$\bot$M5 (3). For $n=2$ it is that of M5-branes along a 1-brane,
M5$\bot$M5 (1). 

\medskip 

We will examine NC versions of these two cases below. In
a commutative limit they should be reduced to M5$\bot$M5 (3) or
M5$\bot$M5 (1), and hence we hereafter assume that 
%%%
$n$ is even.

\subsection{Intersecting NC M5$\bot$NC M5 (3)}\label{sec:M5.M5(3)}

First of all, in order to make our analysis simpler, let us impose 
the following condition, 
\begin{eqnarray}
\Gamma^{A_0A_1A_2}=\Gamma^{B_0B_1B_2}\,, \label{codd}
\end{eqnarray}
in addition to the ansatz (\ref{gluing}). Then the commuting relation
$[M_1,M_2]=0$ requires that
\begin{eqnarray}
[\Gamma^{A_0\cdots A_5},\Gamma^{B_0\cdots B_5}]=0\,, \quad 
[\Gamma^{A_0\cdots A_5},\Gamma^{B_0B_1B_2}]=0\,. \label{4.3}
\end{eqnarray} 
The conditions (\ref{4.3}) are satisfied if $\Gamma^{A_0\cdots A_5}$ and
$\Gamma^{B_0\cdots B_5}$ share even number of indices, and
$\Gamma^{A_0\cdots A_5}$ and $\Gamma^{B_0\cdots B_2}$ share odd number
of indices, respectively. 
%By examining these conditions, we find
%possible 1/4 BPS intersecting NC M5-branes. 
We find 
%%%
two kinds of intersections
%some examples of two types 
under the conditions (\ref{codd}) and
(\ref{4.3}). 
After describing some examples of them,
we move on to more general solutions.
%Then a general solution will be discussed without them. 

\subsubsection*{The first example}

\begin{figure}[ht]
  \begin{center}
    \includegraphics[keepaspectratio=true,height=50mm]{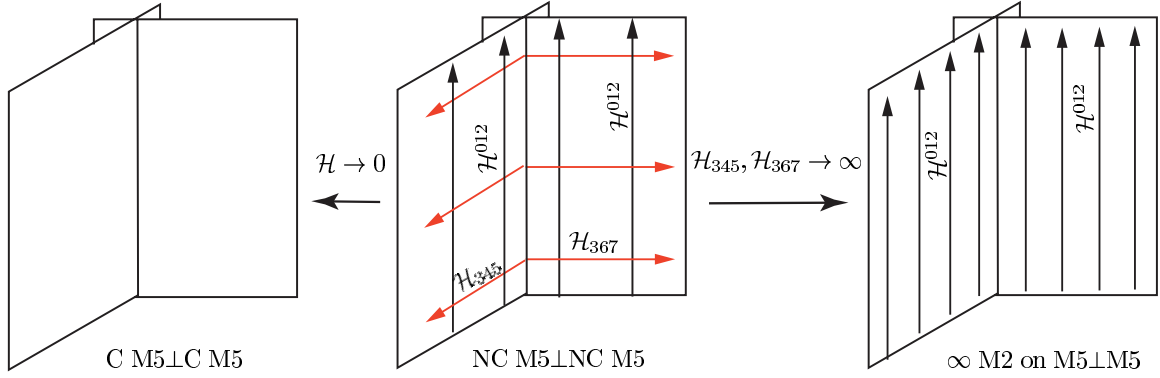}
  \end{center}
  \caption{\footnotesize Strong flux and commutative limits of 
NC M5$\bot$NC M5 (3).  
%NC M5$\bot$NC M5 to M2
} \label{fig:NCM5M5a.eps}
\end{figure}

The first example is M5 (012345)$\bot$M5 (012367) with $\CH^{012}$,
$\CH_{345}$ and $\CH_{367}$\,. To begin with, the vanishing condition of
the $\kappa$-variation surface term is solved for each of the two
M5-branes with two parameters. For the first M5, the gluing matrix is
given by
\begin{eqnarray}
M_1=\e^{\varphi_1\Gamma^{012}}\Gamma^{012345}\,, \quad 
\CH^{012}=-\tanh\varphi_1\,, \quad 
\CH_{345}=\sinh\varphi_1\,, \nonumber 
\end{eqnarray}
and for the second M5 it is  
\begin{eqnarray}
M_2=\e^{\varphi_2\Gamma^{012}}\Gamma^{012367}\,, \quad 
\CH^{012}=-\tanh\varphi_2\,, \quad 
\CH_{367}=\sinh\varphi_2\,. \nonumber 
\end{eqnarray}
From the expression of the electric flux we can read off the 
condition $\varphi_1=\varphi_2$\,, and so 
the gluing matrices and fluxes are given by, respectively, 
\begin{eqnarray}
&&
M_1=\e^{\varphi\Gamma^{012}}\Gamma^{012345}\,, \quad 
M_2=\e^{\varphi\Gamma^{012}}\Gamma^{012367}\,, \label{4.4} \\ 
&& \CH^{012}=-\tanh\varphi\,, \quad 
\CH_{345}=\sinh\varphi\,, \quad 
\CH_{367}=\sinh\varphi\,. \label{4.5}
\end{eqnarray}
%where we choose $\varphi_1=\varphi_2\equiv\varphi$
%since the surface term is deleted by
%$\CH^{012}=\tanh\varphi_1=\tanh\varphi_2$,
%$\CH_{345}=\sinh\varphi_1$ and $\CH_{367}=\sinh\varphi_2$. 
Taking a commutative limit $\varphi\to 0$\,, the configuration described
by (\ref{4.4}) and (\ref{4.5}) reduces to an intersecting M5-branes, M5
(012345)$\bot$M5 (012367) \cite{PT,Tseytlin,GKT1}. 
Taking a strong flux limit $\varphi\to\infty$\,, both fermionic 
boundary conditions reduce to 
\[
\theta=\Gamma^{012}\theta\,, 
\]
so that it should be regarded as a couple of M2-branes in
%%%
  two sets 
of infinitely many M2-branes with $\CH^{012}=-1$ on the 
M5 (012345)$\bot$M5 (012367). 
%Thus in the limit supersymmetry enhances.
These limits are depicted in Fig.\ \ref{fig:NCM5M5a.eps}. 

\vspace*{0.5cm}

\vspace*{0.5cm}

\subsubsection*{The second example}

\begin{figure}[ht]
\vspace*{0.5cm}
  \begin{center}
    \includegraphics[keepaspectratio=true,height=51mm]{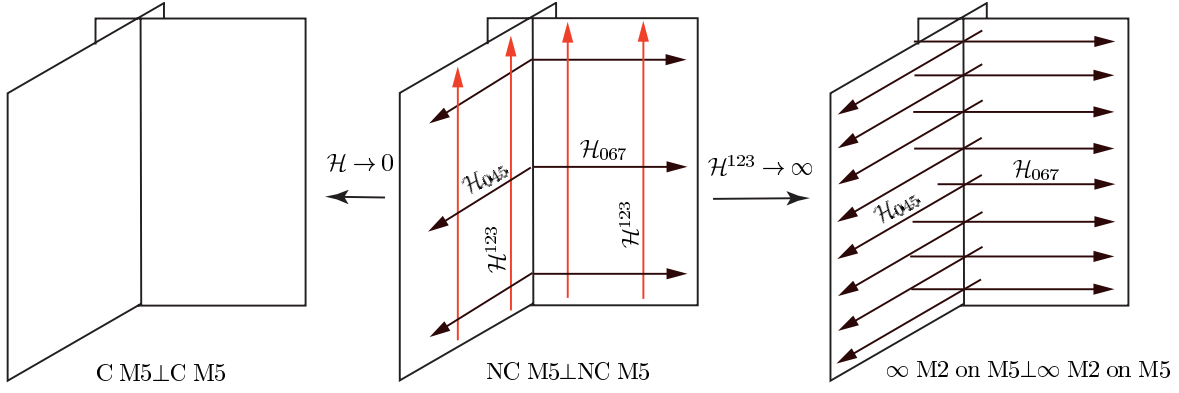}
  \end{center}
  \caption{\footnotesize Strong flux and commutative limits of NC
  M5$\bot$NC M5 (3).}  \label{fig:NCM5M5b.eps}
\end{figure}

The second example is M5 (012345)$\bot$M5 (012367) with $\CH^{123}$,
$\CH_{045}$ and $\CH_{067}$\,.  The construction of projection operators
can be carried out in the similar way and the result is given by
\begin{eqnarray}
&&M_1=\e^{\varphi\Gamma^{123}}\Gamma^{012345}\,, \quad 
M_2=\e^{\varphi\Gamma^{123}}\Gamma^{012367}\,, \nonumber  \\
&& \CH^{123}=\tan\varphi\,, \quad \CH_{045}=
%\sin\varphi\,, \quad 
\CH_{067}=\sin\varphi\,. \nonumber 
\end{eqnarray}
%where we choose $\varphi_1=\varphi_2\equiv\varphi$
%since the surface term disappears
%when
%$\CH^{123}=\tan\varphi_1=\tan\varphi_2$,
%$\CH_{045}=\sin\varphi_1$ and $\CH_{067}=\sin\varphi_2$.
By taking a commutative limit $\varphi\to 0$\,, this configuration
reduces to intersecting M5-branes, M5 (012345)$\bot$M5 (012367). Taking
a strong flux limit $\varphi\to \pi/2$\,, the projection operators
reduce to M2-branes again. But we turned on two different electric
fluxes and so the projection operators are different each other. Hence
we find intersections of two sets of infinitely many M2-branes, M2
(045)$\bot$M2 (067) with $\CH_{045}=\CH_{067}=1$\,. 
%%%
See Fig.\ \ref{fig:NCM5M5b.eps}. 
This sub-brane
system M2$\bot$M2 (0) is nothing but a possible intersecting
configuration of two M2-branes. The corresponding supergravity solution
is found in \cite{PT,GKT1}.
%The latter is nothing but one considered in section \ref{NC M2.M2}.
%See Fig.\ \ref{fig:NCM5M5b.eps}. 
%We examine the boundary condition for
%the NC version of M2$\bot$M2 in Sec.\ 4.3.

\vspace*{0.5cm}

\subsubsection*{The third example}

Finally, by removing the condition (\ref{codd}), we consider the case
with $\Gamma^{A_0A_1A_2}\neq\Gamma^{B_0B_1B_2}$.  In this case,
$[M_1,M_2]=0$ requires that 
\begin{eqnarray}
&& [\Gamma^{A_0\cdots A_5},\Gamma^{B_0\cdots B_5}]=0\,, \qquad 
[\Gamma^{A_0\cdots A_5},\Gamma^{B_3B_4B_5}]=0\,, \qquad \nonumber \\ 
&& [\Gamma^{B_0\cdots B_5},\Gamma^{A_3A_4A_5}]=0\,, \qquad 
[\Gamma^{B_3B_4B_5},\Gamma^{A_3A_4A_5}]=0\,. \nonumber 
\end{eqnarray}  
By examining the above conditions, we shall find intersecting 
M5-branes with four fluxes, 
say, M5 (012345)$\bot$ M5 (012367) with $\CH_{014}$\,,
$\CH_{235}$, $\CH_{026}$ and $\CH_{137}$\,.  

For the first M5-brane, the gluing matrices and fluxes are
\begin{eqnarray}
&&M_1^t=\e^{\varphi_1\Gamma^{235}}\Gamma^{012345}\,, \quad 
\CH_{014}=\sin\varphi_1\,, \quad 
\CH^{235}=\tan\varphi_1\,, \nonumber 
\\\mathrm{or}~~ 
&&M_1^h=\e^{\phi_1\Gamma^{014}}\Gamma^{012345}\,, \quad 
\CH_{235}=\sinh\phi_1\,, \quad 
\CH^{014}=-\tanh\phi_1\,. \nonumber 
\end{eqnarray}
For the second M5-brane those are 
\begin{eqnarray}
&&
M_2^t=\e^{\varphi_2\Gamma^{137}}\Gamma^{012367}\,, \quad 
\CH_{026}=-\sin\varphi_2\,, \quad 
\CH^{137}=\tan\varphi_2\,, \nonumber 
\\
\mathrm{or}~~&&
M_2^h=\e^{\phi_2\Gamma^{026}}\Gamma^{012367}\,, \quad 
\CH_{137}=-\sinh\phi_2\,, \quad 
\CH^{026}=-\tanh\phi_2\,, \nonumber 
\end{eqnarray}
where $0\le \varphi_{1,2}\le\pi/2$ and $0\le \phi_{1,2}<\infty$\,. 
Therefore there are four possibilities,
characterized by boundary conditions with 
\begin{eqnarray}
1) \quad (M_1^t,M_2^t)\,, \qquad 2) \quad (M_1^t,M_2^h)\,, \qquad 
3) \quad (M_1^h,M_2^t)\,, \qquad 4) \quad (M_1^h,M_2^h)\,. \nonumber 
\end{eqnarray}
%$(A_0,A_1,A_2)=(0,1,4)$, $(2,3,5)$
%and $(B_0,B_1,B_2)=(0,2,6)$, $(1,3,7)$ 
All of the configurations reduce to the intersecting M5-branes, 
M5 (012345)$\bot$M5 (012367) in a commutative limit 
$\varphi_{1,2}\to 0$ and $\phi_{1,2}\to 0$\,.

\medskip 

The NC M5-brane described by one of $M_{1,2}^t$ and $M_{1,2}^h$
reduces to infinitely many M2-branes with critical $\CH$ 
on the M5-brane
for $\varphi_{1,2}\to\pi/2$
and $\phi_{1,2}\to\infty$, respectively.
So all of four intersections M5$\bot$M5 reduce to
M2$\bot$M2 with critical fluxes
in the large $\CH$ limit
\begin{eqnarray}
\theta=\Gamma^{014}\theta \quad 
\mbox{and} \quad 
\theta=-\Gamma^{026}\theta\,. \nonumber 
\end{eqnarray}
%which is considered in section \ref{NC M2.M2}.

\medskip 

%The NC version of M2$\bot$M5 
An intersecting configuration M2$\bot$M5 (1) \cite{PT,GKT1} can also be
realized from the NC M5$\bot$M5 obtained above by taking a strong flux
limit. For example, let us consider the NC M5$\bot$M5 characterized by
$(M_1^t,M_2^t)$\,. A NC M5-brane can be regarded as a bound state of M5
and M2, hence we can see M2$\bot$M5 (1) as a sub-brane system of NC
M5$\bot$NC M5. By taking the limit $\varphi_2\to\pi/2$ of the NC
M5$\bot$M5 with $(M_1^t,M_2^t)$, we obtain the gluing matrices for
M2$\bot$M5 (1), 
\begin{eqnarray}
M_1=\e^{\varphi_1\Gamma^{235}}\Gamma^{012345}\, \quad 
M_2=-\Gamma^{026}\,. \nonumber 
%\label{M2.M5:1}
\end{eqnarray}
%characterized by (\ref{M2.M5:2}). 
It should be remarked that the two gluing matrices commute each other
even at this stage.  Further taking $\varphi_1\to\pi/2$\,, we obtain the
intersecting configuration M2 (014)$\bot$M2 (026) as a sub-brane system
on the M5-branes, again.

\medskip 

\begin{figure}[httb]
\vspace*{0.5cm}
  \begin{center}
    \includegraphics[keepaspectratio=true,height=100mm]{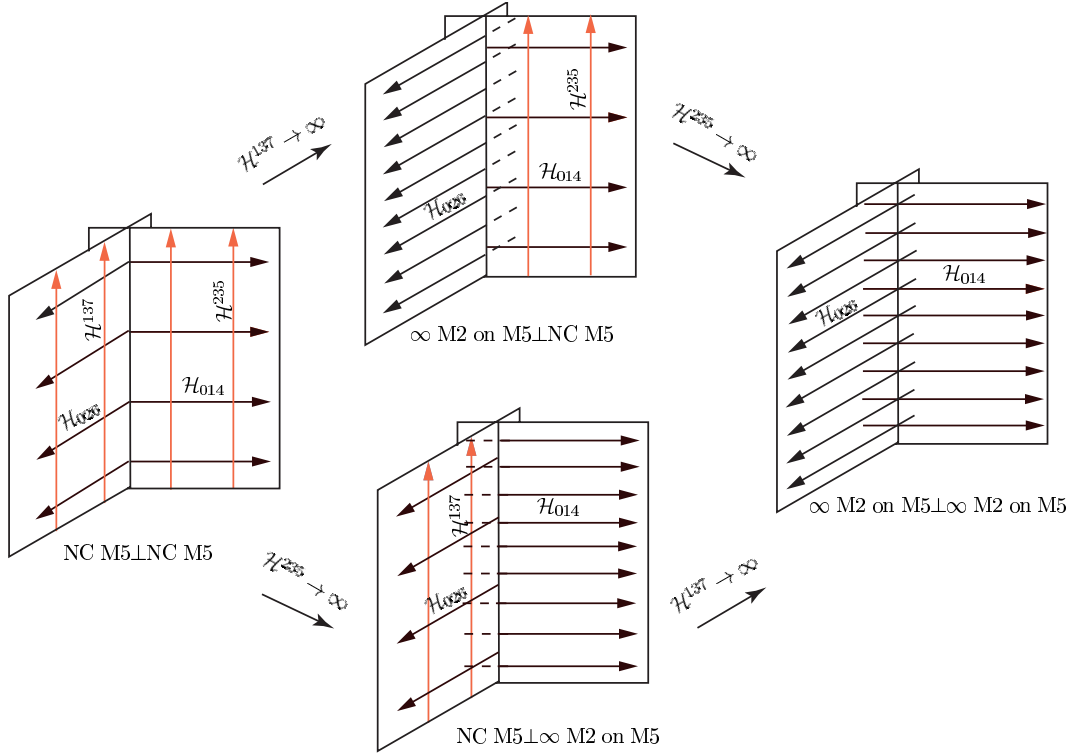}
  \end{center}
  \caption{\footnotesize 
Two sequences of strong flux limits of NC M5$\bot$NC M5 (3).}
  \label{fig:NCM5M5c.eps}
\end{figure}

On the other hand, by taking the limit $\varphi_1\to\pi/2$\,, 
we can obtain another gluing matrices for M2$\bot$M5 (1). 
%it reduces to the intersection of infinitely many M2 (014) with
%$\CH_{014}=1$ on the M5 (012345) and the NC M5 (012367) with $\CH_{026}$
%and $\CH^{137}$ 
\begin{eqnarray}
M_1=\Gamma^{014}\,, \quad 
M_2=\e^{\varphi_2\Gamma^{137}}\Gamma^{012367}\,. \nonumber 
%\label{M2.M5:1}
\end{eqnarray}
These also commute each other. 
%This corresponds to the NC version of M2$\bot$M5. 
Then taking $\varphi_2\to\pi/2$, it reduces to M2 (014)$\bot$M2 (026). These
sequences of the strong flux limits are depicted in Fig.\
\ref{fig:NCM5M5c.eps}.

%We will discuss the boundary conditions for 
%the NC version of M2$\bot$M5 in subsection \ref{NC M2.M5}.

\subsection{Intersecting NC M5$\bot$M5 (1)}\label{sec:M5.M5(1)}

\begin{figure}[httb]
\vspace*{0.5cm}
  \begin{center}
    \includegraphics[keepaspectratio=true,height=55mm]{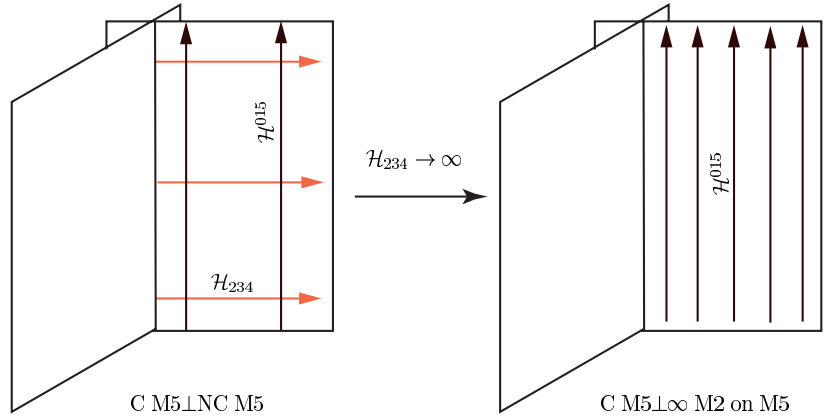}
  \end{center}
  \caption{\footnotesize Strong flux limit of C M5$\bot$NC M5 (1).}
  \label{fig:NCM5M5d.eps}
\end{figure}

Here let us consider 
%%%
configurations of M5$\bot$M5 (1), say
M5 (012345)$\bot$M5 (056789)
 with fluxes 
$\CH_{015}$ and $\CH_{234}$\,.  
There are two kinds of configurations.
%As an example, we will consider . 
The first is characterized by
%an overlap of a NC M5-brane with $\CH^{015}$ and $\CH_{234}$
%and a commutative M5-brane
\begin{eqnarray}
&&
M_1=\e^{\varphi\Gamma^{015}}\Gamma^{012345}\,, \quad 
M_2=\Gamma^{056789}\,,  \quad 
\CH^{015}=-\tanh\varphi\,, \quad \CH_{234}=\sinh\varphi\,, \nonumber 
\end{eqnarray}
and the other is
\begin{eqnarray}
&& M_1=\e^{\varphi\Gamma^{234}}\Gamma^{012345}\,, \quad 
M_2=\Gamma^{056789}\,, \quad 
\CH_{015}=-\sin\varphi\,, \quad \CH^{234}=\tan\varphi\,. \nonumber 
\end{eqnarray}
%with $\CH^{015}=\tanh\varphi$
%and
%$\CH_{234}=\sinh\varphi$.
For $\varphi\to 0$, both reduce to commutative M5$\bot$M5 (1). For a
strong flux limit,  
%the boundary condition with 
$M_1$ reduces to 
$\theta=\pm\Gamma^{015}\theta$\,, 
%for these cases respectively, 
and so it describes an intersection of an M2 (015) on the M5 (012345)
and a commutative M5 (056789). That is, M2$\bot$M5 (1) has been found
again. See Fig \ref{fig:NCM5M5d.eps}.

\section{Summary and Discussions}

We have discussed 1/4 BPS intersecting NC M5-branes from the viewpoint
of $\kappa$-symmetry of a covariant open supermembrane action. We
constructed projection operators for two types of 1/4 BPS intersecting
NC M5-branes. The one is an intersection of two NC M5-branes: NC
M5$\bot$NC M5 (3). The other is that of a NC M5-brane and a commutative
(C) M5-brane: NC M5$\bot$C M5 (1). A NC M5-brane can be viewed as a
bound state of M5 and M2, and the configurations M2$\bot$M5 (1) and
M2$\bot$M2 (0) are realized on the intersecting M5-branes. Taking a
commutative limit the allowed intersecting M5-branes are surely
reproduced: M5$\bot$M5 (3) and M5$\bot$M5 (1). By considering a strong
flux limit, we have found projection operators even 
for M2$\bot$M5 (1) and M2$\bot$M2 (0), which still ensures the
$\kappa$-invariance of the membrane action.

\medskip 

As another task we are going to consider AdS M-branes with constant
three-form fluxes on AdS$_{4/7}\times$S$^{7/4}$ and pp-wave by using the
$\kappa$-symmetry argument (For NC D-branes in flat space and a pp-wave,
see \cite{CH}. Intersecting D-branes in pp-waves are discussed in
\cite{Ohta}.). We hope that it could be reported in the near future
\cite{future}. It is also interesting to consider a $\kappa$-symmetry
argument for M-branes at angle \cite{OT}. 

\medskip 

There are some interesting issues related to open supermembranes, such
as the Poisson structure on the M5-brane \cite{Kawamoto,Rudychev}, the
area-preserving diffeomorphism \cite{Matsuo}, S-duality \cite{Kawano},
and open membrane field theory \cite{HM}. Since the $\kappa$-symmetry of
open supermembrane theory should be closely related to the equation of
motion of M5-brane, it might be interesting to reveal
%%%
the connection between those issues and
the $\kappa$-symmetry argument as a future direction.

\section*{Acknowledgments}

The authors thank the Yukawa Institute for Theoretical Physics at Kyoto
University. Discussions during the 21st Nishinomiya-Yukawa Memorial
Symposium on Theoretical Physics ``Noncommutative Geometry and Quantum
Spacetime in Physics'' were useful to complete this work. They also
thank S.~Terashima for useful comments and helpful discussions.
%The authors thank Seiji Terashima
%for useful comments and helpful discussions.
This work is supported in part by the Grant-in-Aid for Scientific
Research (No.~17540262 and No.~17540091) 
from the Ministry of Education,
Science and Culture, Japan.
The work of K.~Y.\ is supported
in part by JSPS Research Fellowships for Young Scientists.

%\newpage

\end{document}